\documentclass[12pt]{article}
\usepackage{epsfig,amssymb,amsmath,psfrag,subfigure,rotate,color,wasysym}

\allowdisplaybreaks
\newcommand{\insertfig}[2]{\mbox{\epsfxsize=#1cm \epsfbox{#2.eps}}}

\def\XXint#1#2#3{{\setbox0=\hbox{$#1{#2#3}{\int}$ }
\vcenter{\hbox{$#2#3$ }}\kern-.6\wd0}}

\def \be  {\begin{equation}}
\def \ee  {\end{equation}}
\def \ba  {\begin{eqnarray}}
\def \ea  {\end{eqnarray}}
\def \baa {\begin{eqnarray*}}
\def \eaa {\end{eqnarray*}}
\def \lab #1 {\label{#1}}

\newcommand\re[1]{(\ref{#1})}
\def\d{\hbox{{d}\kern-.20em\hbox{l}}}

\def \matrix #1 {\left(\begin{array}{cc} #1 \end{array}\right)}

\def \tr {\mathop{\rm tr}\nolimits}

\def \e  {\mathop{\rm e}\nolimits}

\newcommand{\bit}[1]{\mbox{\boldmath$#1$}}

\newcommand{\ft}[2]{{\textstyle\frac{#1}{#2}}}
\begin{document}

\begin{titlepage}

\thispagestyle{empty}

\vspace*{3cm}

\centerline{\large \bf Null octagon from Deift-Zhou steepest descent}
\vspace*{1cm}

\centerline{\sc A.V.~Belitsky}

\vspace{10mm}

\centerline{\it Department of Physics, Arizona State University}
\centerline{\it Tempe, AZ 85287-1504, USA}

\vspace{2cm}

\centerline{\bf Abstract}

\vspace{5mm}

A special class of four-point correlation functions in the maximally supersymmetric Yang-Mills theory is given by the square of the Fredholm determinant of a generalized Bessel 
kernel. In this note, we re-express its logarithmic derivatives in terms of a two-dimensional Riemann-Hilbert problem. We solve the latter in the null limit making use of the 
Deift-Zhou steepest descent. We reproduce the exact octagonal anomalous dimension in 't Hooft coupling and provide its novel formulation as a convolution of the non-linear 
quasiclassical phase with the Fermi distribution in the limit of the infinite chemical potential.

\end{titlepage}

\setcounter{footnote} 0

\newpage




{\bf 1. Introduction.}  The hexagonalization framework \cite{Basso:2015zoa,Fleury:2016ykk,Eden:2016xvg} for correlation functions in the maximally supersymmetric Yang-Mills theory is based 
upon integrable structures \cite{Beisert:2010jr} of the dual effective two-dimensional theory and provides a truly non-perturbative formalism for their calculation. In generic situations, its use is 
hampered by one's inability to perform exact infinite resummations of all excitations propagating on the world-sheet. However, under a judicial choice of the R-symmetry quantum numbers of 
the operators involved, one can significantly reduce their complexity and furthermore by taking these values to infinity one can suppress the majority of contribution stemming from excitations 
wrapping compact space-like cycles. This in turn implies that corresponding transitions are saturated by the vacuum state and therefore the sums disappear. It was realized some time ago 
\cite{Coronado:2018ypq,Coronado:2018cxj}, that for the particular case of the so-called simplest four-point correlator only two sums remain and moreover their summands factorize such 
that the original observable falls apart into the product of two simpler objects named octagons. With each of them containing just a single sum, the latter was performed in Refs.\ 
\cite{Kostov:2019stn,Kostov:2019auq} where a concise representation was derived in terms of the determinant of a semi-infinite matrix. This result laid out the foundation for the 
analyses in Ref.\ \cite{Belitsky:2019fan,Belitsky:2020qrm,Belitsky:2020qir}, where the octagon was further recast as a Fredholm determinant of an integral operator acting on a semi-infinite 
line. Its kernel was identified as a convolution of the well-known Bessel kernel with a Fermi-like distribution that depends on the external kinematics and the 't Hooft coupling $g$.

In Ref.\  \cite{Belitsky:2019fan}, the null limit of the octagon was addressed. It corresponds to the kinematics where any two nearest-neighbor operators approach a light-like interval such
that one of the conformal cross ratios, i.e., $y$, tends to infinity and all other can be set to zero. Using of the method of differential equations, originally developed by Its-Izergin-Korepin-Slavnov 
for calculations of two-point correlation functions in integrable models, one manages to establish the asymptotic behavior of the octagon as $y\to\infty$ and exactly fix the accompanying coefficients 
as functions of the coupling,
\begin{align}
\label{NullOctagon}
\log \mathbb{O} = - y^2 \frac{\Gamma (g)}{2 \pi^2} + \frac{C (g)}{8} + \dots
\, . 
\end{align}
This Sudakov-like behavior should arise from some kind of a semiclassical limit of the effective two-dimensional path integral. Intuitively this can be observed, making use of the fermionic path 
integral representation of the octagon devised in Ref.\ \cite{Kostov:2019auq,KosPet2020} and subsequent Hubbard-Stratonovich transformation performed to get rid of a nonlocality.
This picture is obscure, however,  within the method employed earlier for the light-cone analysis. Therefore, in the present note an attempt is made to establish a link between the anomalous
dimension $\Gamma (g)$ and semiclassics. Indeed, one finds that with a proper method, $\Gamma(g)$ arises as a WKB phase in the Deift-Zhou non-linear steepest descent method applied
to Riemann-Hilbert problems (RHP) for Fredholm determinants of integrable operators. While, as a result of this consideration, one merely reproduces the function $\Gamma (g)$ at finite coupling with a 
different method, there is a hope  that its complementary novel  representation could shed light on establishing a connection to the same anomalous dimension which arises in a completely different 
context, i.e., the origin limit of the six-gluon scattering amplitude \cite{Basso:2020xts}, which is again governed by a Fredholm determinant akin to the one encountered for the octagon.

{\bf 2. Octagon as a Fredholm determinant.}  According to the results of Refs.\ \cite{Belitsky:2019fan,Belitsky:2020qrm,Belitsky:2020qir}, the octagon is given by the Fredholm determinant
\begin{align}
\label{OtoFredholm}
\mathbb{O} = \det (1 - \mathbb{K} )
\end{align}
of the integral operator
\begin{align}
\mathbb{K} f(x') =\int_{- \infty}^0 dx\, \chi(x) K (x,x') f(x)\,,
\end{align}
with the Bessel kernel \cite{Forrester:1993vtx,Tracy:1993xj}
\begin{align}\notag\label{K-ker}
K (x,x') 
={\sqrt{-x'}\,J_{1}(\sqrt{-x'})J_{0}(\sqrt{-x}) - \sqrt{-x}\,J_{1}(\sqrt{-x})J_{0}(\sqrt{-x'})\over 2(x-x')}\,,
\end{align}
and a cutoff function $\chi (x)$, which depends, in a generic case, on several space-time and R-symmetry cross ratios and the 't Hooft coupling. Compared to the original definition of
Refs.\ \cite{Belitsky:2019fan,Belitsky:2020qrm,Belitsky:2020qir}, the semi-infinite interval was reflected through the origin. This is done in order to have a seamless match to the standard 
conventions adopted for the Bessel model RHP (to be employed below) which is defined on the negative real axis. The focus in this paper will be on the light-like limit when one of the 
variables, namely $y$, tends to infinity with the other set to zero. In this case, the cut-off function significantly simplifies and takes on the form of the Fermi distribution
\begin{align}
\chi (x) = [1 + \e^{(\sqrt{-x} - \mu)/T}]^{-1}
\, ,
\end{align}
where the temperature is set by the 't Hooft coupling $T = 2g$ and the chemical potential is proportional to the cross ratio in question $\mu = 2 g y$.

It is well-known that the Bessel kernel belongs to the class of integrable operators in the nomenclature of Ref.\ \cite{Its:1990} and therefore the generalized Bessel kernel can be cast into 
the form
\begin{align}
\label{GenBesselKer}
\chi(x) K (x,x') = \frac{\bit{f} (x) \cdot \bit{g} (x')}{x - x'}
\, ,
\end{align}
with conveniently chosen two two-vectors $\bit{f}$ and $\bit{g}$ which read
\begin{align}
\bit{f} (x)= \frac{\chi (x)}{2 \pi i}
\left(
{J_0 (\sqrt{-x}) \atop - i \pi \sqrt{-x} J_1 (\sqrt{-x})}
\right)
\, , \qquad
\bit{g} (x) =
\left(
{i \pi \sqrt{-x} J_1 (\sqrt{-x})
\atop
J_0 (\sqrt{-x})  
}
\right)
\, .
\end{align}
These ensure a regular behavior at coincident points $\bit{f} (x) \cdot \bit{g} (x) = 0$, such that $\chi (x)K (x, x) = \bit{f}^\prime (x) \cdot \bit{g} (x) = - \bit{f} (x) \cdot \bit{g}^\prime (x)$.
Notice that the entire dependence on the external parameters $y$ and $g$ is encoded in the vector $\bit{f}$.

The goal is to calculate the Fredholm determinant \re{OtoFredholm}. The crucial observation is that its derivative with respect to external parameters ($\mu$ and $T$,
in the current case) is expressed in terms of the resolvent 
\begin{align}
\mathbb{R} = \frac{\mathbb{K}}{1 - \mathbb{K}} 
\, ,
\end{align}
such that
$\partial \log\det (1 - \mathbb{K}) = - \tr (1 + \mathbb{R}) \partial \mathbb{K}$. In fact, not only the operator $\mathbb{K}$ is integrable but also its resolvent $\mathbb{R}$ as well. 
A simple calculation demonstrates that its kernel can be written as \cite{Its:1990,Harnad:1997hf,DeiftItsZhou97,DeiftIntOper99}
\begin{align}
R (x, x') = \frac{\bit{F} (x) \cdot \bit{G} (x')}{x - x'}
\, ,
\end{align}
with
\begin{align}
\bit{F} (x)
=
\left[ \frac{1}{1 - \mathbb{K}} \bit{f} \right] (x) 
\, , \qquad  
\bit{G} (x)
=
\left[ \frac{1}{1 - \mathbb{K}} \bit{g} \right] (x)
\, .
\end{align}
Thus the analysis boils down to the determination of $\bit{F}$ and $\bit{G}$.

{\bf 3. RHP for the Fredholm determinant.}  It is well known that Fredholm determinants are tau functions associated to RHPs \cite{Tau} and, in particular, coincide with
the Jimbo-Miwa-Ueno conventions \cite{IsoTau} for isomonodromic tau functions of monodromy problems in the theory of ordinary differential equations with rational coefficients. Thus,
the two-component functions $\bit{F}$ and $\bit{G}$ can be determined from a RHP for the two-by-two matrix functions $\bit{Y}$ and $\widetilde{\bit{Y}}$ \cite{Its:1990,DeiftIntOper99} 
(consult also a very comprehensive set of lecture notes \cite{DeiftLectureNotes})
\begin{align}
\bit{F} = \bit{Y}_+ \bit{f}
\, , \qquad
\bit{G} = \widetilde{\bit{Y}}_+ \bit{g}
\, ,
\end{align}
where here and below $\bit{Y}_\pm$ are the limiting values of $\bit{Y}$'s on the real axis, i.e., $\bit{Y}_\pm (x) \equiv \bit{Y}(x \pm i 0)$. Introduce the jump matrices such that they possess 
the components $\{ \bit{V} \}_{ij} = V_{ij}$ and $\{ \widetilde{\bit{V}} \}_{ij} = \widetilde{V}_{ij}$ with
\begin{align}
V_{jk} (x) = \delta_{jk} - 2 \pi i f_j (x) g_k (x)
\, , \qquad
\widetilde{V}_{jk} (x) = \delta_{jk} + 2 \pi i g_j (x) f_k (x) 
\, ,
\end{align}
for $x \in \mathbb{R}_- \equiv (-\infty, 0]$ and $\delta_{jk} $ elsewhere\footnote{This implies that the extension of the cut-off function $\chi$ from $\mathbb{R}_-$ to the entire real axis is
$\chi (x) = \{ [1 + \e^{(\sqrt{-x} - \mu)/T}]^{-1} \, , x \leq 0 ; 0 \, , x > 0 \}$.}. Observing that $\widetilde{\bit{V}} = (\bit{V}^{-1})^T$, the matrices $\bit{Y}$ and 
$\widetilde{\bit{Y}}$ are related to each other by the very same relation and thus it suffices to determine only one RHP function since
\begin{align}
\widetilde{\bit{Y}} = (\bit{Y}^{-1})^T
\, .
\end{align}
The RHP that one has to solve is as follows:
\begin{align}
\label{YRHP}
\begin{array}{ll}
\bit{Y}(z) &\mbox{is analytic in $\mathbb{C}\backslash \mathbb{R}_-$} \, ,
\\
\bit{Y}_+ (z) = \bit{Y}_- (z) \bit{V} (z)  
&\mbox{for $z \in  \mathbb{R}_-$} \, ,
\\
\bit{Y} (z) \stackrel{z \to \infty}{\to} \bit{1} \, . 
&
\end{array}
\end{align}

Having found $\bit{Y}$, one can immediately calculate the Fredholm determinant in its terms. Namely, making use of the result of Ref.\ \cite{DetasRHPproof}, the derivative $\partial$
of the determinant with respect to its parameters $\mu$ or $T$ is
\begin{align}
\label{DetIntKernel}
\partial \log\det (1- \mathbb{K}) 
=
\omega
-
\int_{\mathbb{R_-}} d x \, \partial [ \bit{f}' \cdot \bit{g}] (x)
+
\ft12
\int_{\mathbb{R_-}}  d x  \tr [\bit{V}^\prime \bit{V}^{-1} (\partial \bit{V}) \bit{V}^{-1}] (x)
\, ,
\end{align}
where the Jimbo-Miwa-Ueno differential reads
\begin{align}
\omega \equiv 
\int_{\mathbb{R_-}} \frac{d x}{2 \pi i}
\tr [\bit{Y}_+^{-1} \bit{Y}_+^\prime (\partial \bit{V}) \bit{V}^{-1} ] (x)
\, .
\end{align}
In these formulas the prime on all symbols designates differentiation with respect to the spectral parameter $x$, e.g., $\bit{Y}' (x) = \partial_x \bit{Y} (x)$.

{\bf 4. Dressing procedure.} The jump matrices for the RHP \re{YRHP} are very complicated since they depend on the Bessel functions. Therefore, one has to simplify $\bit{V}$ 
first if one hopes to solve the problem. This can be achieved by a standard dressing procedure which allows one to completely eliminate any reference to the ubiquitous Bessel 
functions. This construction was recently used in the generalized Airy problem in Ref.\ \cite{ModAiryRHP} which is akin to the one studied in this work. Presently, this technique is 
applied to the current circumstances making necessary adjustments along the way.

There is some freedom in the choice of the dressing function. The RHP for the Bessel model, which is recalled in the Appendix A, is not general enough for the purpose. One can 
introduce an extra parameter $z_0$ through the shift of the Bessel model RHP contour to the left and thus construct a modified Bessel model solution $\widehat{\bf\Psi}$. Employing 
the latter, one devises the dressed $\bit{Y}$ via
\begin{align}
\label{DressedY}
\begin{array}{ll}
{\bf\Phi}_{\rm I, VI} \!\!
&
= \bit{Y} \widehat{\bf \Psi}_{\rm I, VI}
\, , \\
{\bf \Phi}_{\rm II} \!\!
&
= \bit{Y} \widehat{\bf \Psi}_{\rm II}
\left(
\begin{array}{cc}
1 & 0 \\
- \frac{\chi}{1 - \chi} & 1
\end{array}
\right)
\, , \\
{\bf \Phi}_{\rm III} \!\!
&
= \bit{Y} \widehat{\bf \Psi}_{\rm III}
\left(
\begin{array}{cc}
1 & 0 \\
\frac{\chi}{1 - \chi} & 1
\end{array}
\right)
\, ,
\end{array}
\end{align}
for the four regions partitioned by the modified Bessel model RHP contour $C = C_1 \cup C_2 \cup C_3 \cup C_4$ as explained in the Appendix A. The additional right-most matrices 
in Eq.\ \re{DressedY} depending on $\chi$ emerge from the factorization of the jump matrix on $C_2$ for the bare product $\bit{Y} \widehat{\bf \Psi}$ in terms of lower-triangular and 
anti-diagonal matrices. The dressed functions obey the jump relations
\begin{align}
\label{dressedRHP}
\begin{array}{ll}
{\bf \Phi}_+ 
= {\bf \Phi}_- 
\left(
\begin{array}{cc}
0 & 1 - \chi \\
- \frac{1}{1 - \chi} & 0
\end{array}
\right)
\, , 
&
\mbox{for} \ z\in C_2
\\
{\bf \Phi}_+ 
= {\bf \Phi}_- 
\left(
\begin{array}{cc}
1 & 0 \\
\frac{1}{1 - \chi} & 1
\end{array}
\right)
\, , 
&
\mbox{for} \ z\in C_{1,3}
\\
{\bf \Phi}_+ 
= {\bf \Phi}_- 
\left(
\begin{array}{cc}
1 & 1 - \chi \\
0 & 1
\end{array}
\right)
\, , 
&
\mbox{for} \ z\in C_4
\end{array}
\end{align}
These can be deduced from the jump matrices for $\bit{Y}$ and $\widehat{\bf\Psi}$ together with the following instrumental relation between the two-dimensional
vectors $\bit{f}$ and $\bit{g}$ and the solutions to the modified Bessel model RHP \re{ModBesModelRHP},
\begin{align}
2 \pi i \chi^{-1} (z)
\bit{f} (z) 
&
= 
\left\{
\begin{array}{cc}
z> z_0 :
& 
\widehat{\bf \Psi}_+ (z) \left({1 \atop 0} \right)
\\[2mm]
z< z_0 :
& 
\widehat{\bf \Psi}_+ (z) \left({1 \atop 1} \right)
\end{array}
\right.
\, , \\
\bit{g}^T (z) 
&
= 
\left\{
\begin{array}{cc}
z> z_0 :
& 
(0,1) \widehat{\bf \Psi}_+^{-1} (z)
\\[2mm]
z< z_0 :
& 
(-1, 1) \widehat{\bf \Psi}_+^{-1} (z) 
\end{array}
\right.
\, .
\end{align}
The uniqueness of the solution to the RHP \re{dressedRHP} follows a standard argument and will not be repeated here, see, e.g., \cite{DeiftItsZhou97}.

Making use of the above definitions, the Fredholm determinant can be rewritten in terms of the $21$-matrix element of the dressed RHP solution
\begin{align}
\label{LogDetFinal}
\partial \log\det(1 - \mathbb{K})
=
-
\int_{\mathbb{R}_-}
\frac{d z}{2 \pi i} (\partial \chi) (\widehat{\bf\Phi}_+^{- 1} \widehat{\bf\Phi}'_+ )_{21}
\, ,
\end{align}
where
\begin{align}
\widehat{\bf \Phi}_+
= 
\left\{
\begin{array}{ll}
z> z_0 :
& 
{\bf \Phi}_{\rm I}
\, ,
\\[2mm]
z< z_0 :
& 
{\bf \Phi}_{\rm II}
\left(
\begin{array}{cc}
1 & 0
\\
\frac{1}{1 - \chi} &  1\end{array}
\right)
\, .
\end{array}
\right.
\end{align}

{\bf 5. Deift-Zhou steepest descent.} As one is interested in the leading asymptotics of the Fredholm determinant for $\mu \to \infty$, it is instructive to rescale the complex variable
$z$ by $\mu^2$, i.e., $z \to \mu^2 z$, such that the cut-off depends on the ratio $\mu/T$, i.e., $\chi (z) \to \chi (z) = [1 + \e^{(\sqrt{-z} -1) \mu/T}]^{-1}$. However, as one is taking
the limit in question, one immediately faces a predicament since the jump matrices for ${\bf\Phi}$ (see Eqs.\ \re{dressedRHP}) are not decaying to constant matrices. Here is where 
the nonlinear steepest descent comes to the rescue. The idea of the Deift-Zhou method \cite{DeiftZhou94} (see also sect.\ 4 of \cite{DeiftStrongOrthPol} for a concise summary) is 
to perform a set of successive transformations of the original RHP and simplify the problem at each step while determining contributions to the asymptotic solution. The goal is to 
reduce the problem to jump matrices which exponentially approach the ones with constant matrix elements. The ultimate RHP is a small-norm RHP for a function $\bit{R}$ with 
asymptotically unit jumps and normalized behavior at infinity. The analysis is thus divided in the sequence
\begin{align}
\label{DZhChain}
{\bf \Phi} \to \bit{T} \to \bit{R}
\, ,
\end{align}
where in the first step one rotates the ``phases'' away by constructing a function $g$ which is the analogue of the WKB phase in the linear quasiclassical theory and then in the second 
step, one finds approximate solutions to the RHP for $\bit{T}$, which are known as parametrices $\bit{P}$. The latter asymptotically yield the small-norm RHP for $\bit{R} = 
\bit{T} \bit{P}^{-1}$.

{\bf 5. First transformation.} In the first step of the Deift-Zhou chain, one define the matrix function $\bit{T}$ by eliminating the exponential growth of the jump matrices as $\mu \to \infty$ 
by introducing a function $g$ as follows
\begin{align}
\bit{T} = \e^{ - \mu ( V_0/2 ) \sigma_3} {\bf\Phi} \e^{\mu ( g +  V_0/2 ) \sigma_3} 
\, ,
\end{align}
where 
\begin{align}
V (z) = \frac{1}{\mu} \log (1 - \chi (z))
\, ,
\end{align}
and we shifted the origin to $z_0$, such that $V_0 \equiv V (z_0)$. The function $g$ is yet to be determined and is the main focus of a successful RHP analysis. For the transformed 
$\bit{T}$, the jump matrices are
\begin{align}
\label{RHPforT}
\begin{array}{ll}
\bit{T}_+ 
= \bit{T}_- 
\left(
\begin{array}{cc}
0 & \e^{-\mu(g_+ + g_- - V + V_0)}
\\
- \e^{\mu (g_+ + g_- - V + V_0)} & 0
\end{array}
\right)
\, , 
&
\mbox{for} \ z\in C_2
\\
\bit{T}_+ 
= \bit{T}_- 
\left(
\begin{array}{cc}
1 & 0
\\
\e^{\mu (2 g - V + V_0)} & 1
\end{array}
\right)\, , 
&
\mbox{for} \ z\in C_{1,3}
\\
\bit{T}_+ 
= \bit{T}_- 
\left(
\begin{array}{cc}
1 & \e^{- \mu (2 g - V + V_0)}
\\
0 & 1
\end{array}
\right)\, , 
&
\mbox{for} \ z\in C_{4}
\end{array}
\, .
\end{align}
A natural way to choose the function $g$ is to impose the condition that $g$ solves the RHP
\begin{align}
\label{gRHP}
g_+ + g_- - V + V_0 = 0
\, ,
\end{align}
for $z<z_0$ such that one gets a constant jump matrix for the $C_2$ portion of the contour
\begin{align}
\bit{T}_+ 
= \bit{T}_- 
\left(
\begin{array}{cc}
0 & 1
\\
- 1 & 0
\end{array}
\right)
\, , 
\qquad
\mbox{for} \ z\in C_2
\, ,
\end{align}
with the asymptotic behavior being
\begin{align}
\bit{T} (z) \stackrel{z \to \infty}{\to} z^{\sigma_3/4} A
\, .
\end{align}

{\bf 6. Deift-Zhou phase.} The phase function is found from the RHP \re{gRHP} subject to the asymptotic condition
\begin{align}
g(z) \stackrel{z\to\infty}{=} - \sqrt{z} + c + O (1/\sqrt{z})
\end{align}
which is imposed to cancel the untamed behavior of the modified Bessel model RHP solution \re{BesselModelAsyBevavior}. It is easier to solve the RHP for $g'$ first and then 
integrate it back to $g$. Differentiating both sides of \re{gRHP} with respect to the spectral parameter $z$, one deduces
\begin{align}
g'_+ (z) + g'_- (z) = - \frac{1}{2 T} \frac{\chi(z)}{\sqrt{-z}}
\, ,
\end{align}
with the asymptotic behavior at infinity
\begin{align}
g' (z) \stackrel{z \to \infty}{=} - \frac{1}{2 \sqrt{z}} + O (z^{-3/2})
\, .
\end{align}
The solution is easily found to be
\begin{align}
g' (z) = - \frac{1}{2 \sqrt{z - z_0}}
\left[
1 + \frac{1}{T}
\int_{- \infty}^{z_0}
\frac{d z'}{2 \pi} \sqrt{1 - \frac{z_0}{z}} \frac{\chi (z')}{z' - z}
\right]
\, ,
\end{align}
while integrating this result, one concludes that
\begin{align}
g (z) = - \sqrt{z - z_0} + \frac{1}{2 T}
\int_{- \infty}^{z_0}
\frac{d z'}{2 \pi i} \frac{\chi (z')}{\sqrt{- z'}} \log \frac{\sqrt{z_0 - z'} + i \sqrt{z - z_0}}{\sqrt{z_0 - z'} - i \sqrt{z - z_0}}
\, .
\end{align}
In fact, these integrals can be performed exactly for the limiting Fermi distribution as $\mu\to\infty$. Keeping only the leading term, one substitutes $\chi$ by the unit step
\begin{align}
\chi_\infty (z) = \theta (-1 \leq z \leq 0)
\, .
\end{align}
and explicitly obtains
\begin{align}
g_\infty' (z) 
= - \frac{1}{2 \sqrt{z - z_0}}
&
+  \frac{i}{4 \pi T}
\frac{1}{\sqrt{-z}}
\log 
\frac{ \sqrt{z-z_0} - \sqrt{z (z_0+1)} 
}{
 \sqrt{z-z_0} + \sqrt{z (z_0+1)}
}
\nonumber\\
&
-
 \frac{i}{4 \pi T}
\frac{1}{\sqrt{z_0-z}}
\log \frac{1-\sqrt{z_0+1}}{1 + \sqrt{z_0+1}}
\, , \\
g_\infty (z) 
= 
 c_{1/2}  \sqrt{z-z_0}
&
+
\frac{\sqrt{-z}}{2 \pi T} \log \frac{\sqrt{z - z_0} - \sqrt{z (1 + z_0)}}{\sqrt{z - z_0} + \sqrt{z (1 + z_0)}}
\nonumber\\
&
-
\frac{i}{\pi T} \log \frac{\sqrt{1 + z_0} + i \sqrt{z - z_0}}{\sqrt{1 + z}}
-
\frac{\sqrt{-z_0}}{2 T}
\, ,
\end{align}
where the coefficient in front of $\sqrt{z-z_0}$ will be of importance in what follows
\begin{align}
\label{c12}
 c_{1/2} 
 =
- 1 +\frac{1}{\pi T} \log\frac{1 + \sqrt{1 + z_0}}{\sqrt{-z_0}} 
\, .
\end{align}

Making use of the explicit solution for $g(z)$, one can immediately verify that the jump matrices for the $C_{1,3,4}$ portions of the RHP
contour $C$ approach the sought after unit matrices everywhere in the complex plane except for a small neighborhood around $z = z_0$. 
On the interval $z \in (-\infty, z_0)$ the jump matrix is purely off-diagonal with constant elements. Thus, the approximate solution to the problem 
is determined by first ignoring the small off-diagonal jumps and finding the so-called global parametrix $\bit{P}_\infty$. Then zooming into 
the $\epsilon$-vicinity of $z_0$, where the point-wise in $z$ convergence of the jump matrices breaks down, one determines the local parametrix 
$\bit{P}_{\rm loc}$.

{\bf 7. Global parametrix.}  The global parametrix is found from the following RHP: $\bit{P}_\infty$ is analytic in $\mathbb{C} \backslash (-\infty, z_0)$
and obeys the jump condition  
\begin{align}
\bit{P}_{\infty,+} 
= \bit{P}_{\infty,-} 
\left(
\begin{array}{cc}
0 & 1
\\
- 1 & 0
\end{array}
\right)
\, , 
\qquad
\mbox{for} \ z\in (-\infty, z_0)
\, ,
\end{align}
with
\begin{align}
\bit{P}_\infty (z) \stackrel{z \to \infty}{\to} z^{\sigma_3/4} A
\, .
\end{align}
One can immediately verify that the solution is
\begin{align}
\bit{P}_\infty (z) = (z - z_0)^{\sigma_3/4} A
\, .
\end{align}

\begin{figure}[t]
\begin{center}
\mbox{
\begin{picture}(0,100)(180,0)
\put(0,0){\insertfig{13}{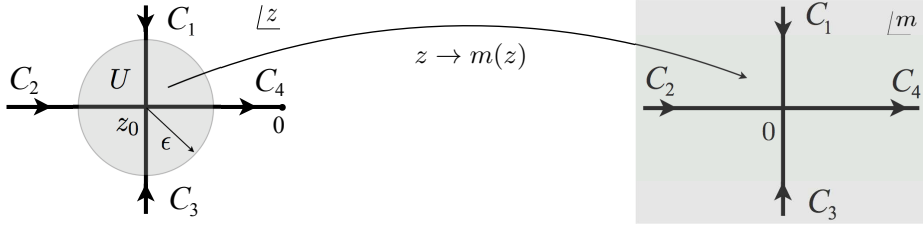}}
\end{picture}
}
\end{center}
\caption{\label{LocalParametrixPic} The $\epsilon$-vicinity of $z_0$ defines the disk $U$. The map $m$ \re{Mapm} transforms the disk into the entire complex plane.}
\end{figure}

{\bf 8. Local parametrix.} In the vicinity of $z_0$, the combination entering the exponent of the off-diagonal elements of the jump matrices for $\bit{T}$ is
\begin{align}
\left( g - \ft12 (V - V_0) \right)_\infty
\stackrel{z \to z_0}{=} 
(z - z_0)^{1/2} \left[ c_{1/2} + c_{3/2} (z - z_0)^2 + \dots \right]
\, ,
\end{align}
with $c_{1/2}$ as determined above in Eq.\ \re{c12} and
\begin{align}
c_{3/2}
=
- \frac{1}{3 \pi T z_0 \sqrt{1 + z_0}}
\, .
\end{align}
So it vanishes at least as a square root such that in the $\epsilon$-vicinity of $z_0$, the jump matrices $\widehat{\bit{M}}$ become
\begin{align}
\label{AiryJumps}
\widehat{\bit M}_{C_1, C_3} = 
\left(
\begin{array}{cc}
1 & 0 \\
1 & 1
\end{array}
\right)
\, ,
\qquad
\widehat{\bit M}_{C_2} = 
\left(
\begin{array}{cc}
0 & 1 \\
-1 & 0
\end{array}
\right)
\, ,
\qquad
\widehat{\bit M}_{C_4} = 
\left(
\begin{array}{cc}
1 & 1 \\
0 & 1
\end{array}
\right)
\, .
\end{align}
They are the same as in the Airy model reviewed in the Appendix B.  Therefore, when one seeks a conformal map from the small disk $U$ of $z_0$ 
\begin{align}
\label{Mapm}
z \to m (z)
\, , \qquad
\ft{2}{3} m^{3/2} (z) =  g - \ft12 (V - V_0)
\end{align}
to the entire complex plane $\mathbb{C}$ (see Fig.\ \ref{LocalParametrixPic}), one can reconcile the asymptotic behavior of the nonlinear steepest descent phase with 
the form of the jump matrices \re{AiryJumps} provided one imposes the condition 
\begin{align}
\label{z0ToT}
c_{1/2} = 0
\, . 
\end{align} 
Then
\begin{align}
\left( g - \ft12 (V - V_0) \right)_\infty
\stackrel{z \to z_0}{=} 
c_{3/2} (z - z_0)^{3/2} + \dots 
\, , 
\end{align}
as expected for the Airy model RHP \re{AiryModelAsy}. 

The RHP for the local parametrix $\bit{P}_{\rm loc}$ thus reads
\begin{align}
\bit{P}_{\rm loc, +} = \bit{P}_{\rm loc, -} \widehat{\bit M}
\, ,
\end{align}
and $\bit{P}_{\rm loc} \to \bit{P}_{\rm \infty}$ as $\mu \to \infty$ on the boundary $\partial U$. This local parametrix is expressed as
\begin{align}
\label{LocalAiryParametrix}
\bit{T}^{\rm loc} (z) = \left( \frac{z - z_0}{\mu^{2/3} m (z)} \right)^{\sigma_3/4} {\bit\phi} \left(\mu^{2/3} m (z) \right) \e^{\mu \left(  g - (V - V_0)/2 \right) \sigma_3}
\, .
\end{align}
by means of the solution $\bit{\phi} (z)$ to the Airy model RHP \re{AirySol1} -- \re{AirySol4}.

{\bf 9. Small-norm RHP.} Having found the global and local parametrices it is easy to show that
\begin{align}
\bit{R}
=
\left\{
\begin{array}{ll}
\bit{T} \bit{P}_{\rm loc}
&
\mbox{for}\ z \in U
\, ,
\\[2mm]
\bit{T} \bit{P}_{\infty}
& 
\mbox{for}\ z \in \mathbb{C}\backslash U
\, ,
\end{array}
\right.
\end{align}
obeys a small-norm RHP with the asymptotic solution $\bit{R} \to \bit{1}$ as $\mu \to \infty$. Having reached this conclusion, one can invert the set of transformations \re{DZhChain}
and determine the asymptotic behavior of ${\bf\Phi}$ that will feed into the Fredholm determinant \re{LogDetFinal}.

{\bf 10. Asymptotic behavior of the determinant.} With all parametrices in hand, one can finally extract the asymptotic form for the Fredholm determinant of the
generalized Bessel kernel \re{GenBesselKer}. In terms of the RHP matrix $\bit{T}$, the derivative with respect to the chemical potential $\mu$ reads
\begin{align}
\label{determinant}
\partial_\mu \log\det (1 - \mathbb{K})
=
- \frac{1}{T}
\int_{\mathbb{R}_-}
\frac{dz}{2 \pi i} \chi (z)
(
\widehat{\bit{T}}{}^{-1}
\widehat{\bit{T}}{}^{\prime}
)_{21}
\e^{-\mu (2 g_+ - V + V_0)}
\, , 
\end{align}
where
\begin{align}
\label{ThatDef}
\widehat{\bit{T}} (z)
= 
\left\{
\begin{array}{cl}
z> z_0 :
& 
\bit{T}_+
(z)
\\[2mm]
z< z_0 :
& 
\bit{T}_+ (z) 
\left(
\begin{array}{cc}
1 & 0
\\
\e^{\mu (2 g_+ - V + V_0)}&  1
\end{array}
\right)
\end{array}
\right.
\, .
\end{align}
The contributions to the integral from the global and local parametrices is found by splitting the real negative axis into three regions $\mathbb{R}_- = (-\infty, z_0-\epsilon) 
\cup [z_0 - \epsilon, z_0 + \epsilon] \cup (z_0 + \epsilon, 0]$ and calculating the former in the limit $\mu \to \infty$.

For $z \in (- \infty, z_0)$, one finds
\begin{align}
(
\widehat{\bit{P}}^{-1}_\infty
\widehat{\bit{P}}^{\prime}_\infty
)_{21}
&=
\mu
(2 g' - V')
\e^{\mu (2 g_+ - V + V_0)} 
\nonumber\\
&- \frac{i}{2 (z - z_0) }
\e^{\mu (2 g_+ - V + V_0)} 
\cos
\left(
\mu (2 g_+ - V + V_0) 
\right)
\, .
\end{align}
The first term is dominant as $\mu \to \infty$ and its contribution to Eq.\ \re{determinant} evaluates to
\begin{align}
\label{DZhintegral}
\int_{-1}^{z_0}
\frac{dz}{2 \pi i} (2 g' - V')_\infty =
\frac{\sqrt{1 + z_0}}{\pi}
+\frac{1}{\pi^2 T}
\left[
\sqrt{1 + z_0} \log\frac{1 + \sqrt{1 + z_0}}{\sqrt{-z_0}} - \log \sqrt{-z_0}
\right]
\, .
\end{align}

For $z \in [z_0 - \epsilon, z_0 + \epsilon]$, one immediately finds using Eq.\ \re{LocalAiryParametrix},
\begin{align}
(
\widehat{\bit{P}}_{\rm loc}^{-1}
\widehat{\bit{P}}_{\rm loc}^{\prime}
)_{21}
= \mu^{2/3} m^\prime (z) \left( \bit{\phi}^{-1}_+ \bit{\phi}^\prime_+ \right)_{21}  (\mu^{2/3} m (z))  \e^{\mu (2 g_+ - V + V_0)}
\, ,
\end{align}
where the matrix element evaluates to
\begin{align}
\left( \bit{\phi}^{-1}_+ \bit{\phi}^\prime_+ \right)_{21} (z) = i K_{\rm Ai} (z, z)
\, ,
\end{align}
and is expressed via the Airy kernel at coincident points
\begin{align}
K_{\rm Ai} (z, z') = \frac{{\rm Ai} (z) {\rm Ai}^\prime (z^\prime) - {\rm Ai} (z^\prime) {\rm Ai}^\prime (z) }{z - z^\prime}
\, .
\end{align}
Cumulatively,
\begin{align}
(
\widehat{\bit{P}}_{\rm loc}^{-1}
\widehat{\bit{P}}_{\rm loc}^{\prime}
)_{21}
= i \mu^{2/3} m^\prime (z) K_{\rm Ai} \left(s^{2/3} m(z), s^{2/3} m(z) \right)  \e^{\mu (2 g_+ - V + V_0)}
\, ,
\end{align}
and yields a subleading effect compared to the contribution of the global parametrix to the determinant.

Finally, for $z \in (z_0 + \epsilon, 0]$, making use of the definition \re{ThatDef}, one deduces that
\begin{align}
(
\widehat{\bit{T}}^{-1}
\widehat{\bit{T}}^{\prime}
)_{21}
\simeq
(
\widehat{\bit{P}}_{\infty}^{-1}
\widehat{\bit{P}}_{\infty}^{\prime}
)_{21}
=
\frac{i}{4 (z - z_0)} + \dots
\, .
\end{align}
Taking into account the behavior of the exponential factor in the integrand of Eq.\ \re{determinant} as $\mu \to \infty$, the evaluation of the integral results again in a power-suppressed 
effect compared to the  $(-\infty, z_0)$-region. Thus, it does not, similarly to the local parametrix, affect the leading scaling behavior of the Fredholm determinant and, therefore, can 
be ignored.

As was just determined, the leading asymptotics arises only from the global parametrix in the region $(-\infty, z_0)$ and is given by the convolution of the (derivative of the) Deift-Zhou 
phase and the Fermi distribution. The result is a function of the parameter $z_0$. The latter is in turn is related to the temperature $T$ by solving Eq.\ \re{z0ToT}, which stems
from the sewing condition between the global and local parametrices at the disk boundary $\partial U$. Employing Eq.\ \re{c12}, it can be inverted and one finds
\begin{align}
\label{z0Tog}
z_0 = - \frac{1}{\cosh^2 (\pi T)}
\, .
\end{align}
Its substitution to Eq.\ \re{DZhintegral} yields the $\mu$ derivative of the Fredholm determinant
\begin{align}
\partial_\mu \log\det (1 - \mathbb{K})
\stackrel{\mu \to \infty}{=}
- \frac{\mu}{T}  \int_{-1}^{z_0} \frac{dz}{2 \pi i} (2 g' - V')_\infty = - \frac{\mu}{(\pi T)^2} \log\cosh(\pi T)
\, .
\end{align}
Recalling that $T = 2 g$ and $\mu = 2 g y$ and integrating both sides of the above equation with respect to $y$, one uncovers the leading $y^2$-behavior in Eq.\ \re{NullOctagon}.

\vspace{0.4cm}

{\bf Acknowledgments.} We would like to thank Gregory Korchemsky for collaboration on the project and useful discussions as well as careful reading of the manuscript
and instructive comments. This research was supported by the U.S. National Science Foundation under the grant PHY-1713125.


\vspace{0.4cm}

\appendix

\begin{figure}[t]
\begin{center}
\mbox{
\begin{picture}(0,110)(230,0)
\put(0,-15){\insertfig{16}{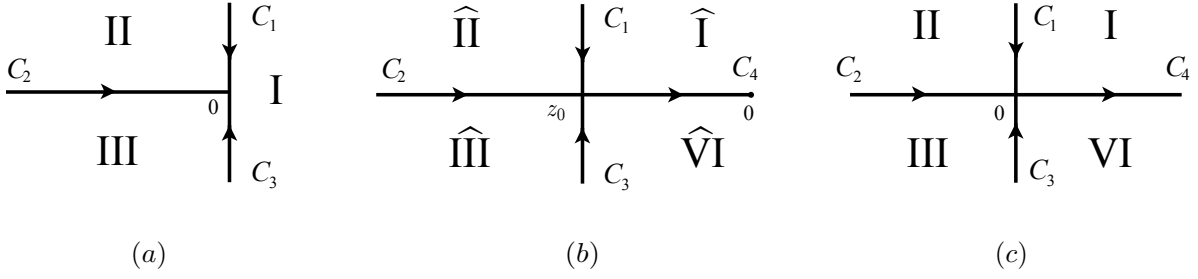}}
\end{picture}
}
\end{center}
\caption{\label{ModelRHPpics} RHP contours for the Bessel (a), modified Bessel (b) and Airy (c) model RHPs.}
\end{figure}

{\bf A. (Modified) Bessel model RHP.} The solution to the Bessel model RHP is analytic everywhere in the complex plane except on the contour 
$C = C_1 \cup C_2 \cup C_3$ shown in the  panel $(a)$ of Fig.\ \ref{ModelRHPpics}, where the former develops jumps
\begin{align}
{\bf\Psi}_+ (z) = {\bf\Psi}_- (z) \bit{M}
\, , \qquad\mbox{for}\qquad z \in C
\, ,
\end{align}
with corresponding jump matrices
\begin{align}
\bit{M}_{C_1, C_3} = 
\left(
\begin{array}{cc}
1 & 0 \\
1 & 1
\end{array}
\right)
\, ,
\qquad
\bit{M}_{C_2} = 
\left(
\begin{array}{cc}
0 & 1 \\
-1 & 0
\end{array}
\right)
\, ,
\end{align}
and possesses a uniform asymptotic behavior
\begin{align}
\label{BesselModelAsyBevavior}
{\bf\Psi}(z) \stackrel{z \to \infty}{\to} (\pi \sqrt{z})^{- \sigma_3/2} A \e^{\sqrt{z} \sigma_3}
\, ,
\end{align}
where $A \equiv (1+i\sigma_1)/\sqrt{2}$ and the Pauli matrices $\sigma_1$ and $\sigma_3$. The solution can be written concisely as \cite{BesselRHPcite}
\begin{align}
{\bf\Psi} (z)
&
=
\left(
\begin{array}{cc}
I_0 (\sqrt{z})  
& 
\ft{i}{\pi} K_0 (\sqrt{z})  
\\
i \pi \sqrt{z} I_1 (\sqrt{z})  
& 
\sqrt{z} K_1 (\sqrt{z})  
\end{array}
\right)
H (0, z)
\, , \\
\end{align}
in terms of the modified Bessel functions of the first and second kind with
\begin{align}
H_{\rm I} (0, z)
=
\left(
\begin{array}{cc}
1 & 0 \\
0 & 1
\end{array}
\right)
\, , \qquad
H_{\rm II} (0, z)
=
\left(
\begin{array}{cc}
1 & 0 \\
-1 & 1
\end{array}
\right)
\, , \qquad
H_{\rm III} (0, z)
=
\left(
\begin{array}{cc}
1 & 0 \\
1 & 1
\end{array}
\right)
\, ,
\end{align}
corresponding to the regions I, II and III, respectively.

What one needs in the dressing procedure used in the body of the paper is the solution to the modified Bessel model RHP. Namely, shifting the imaginary portion of the contour to the left
by a real negative parameter $z_0$, one ends up with a contour shown in the middle panel $(b)$ of Fig.\ \ref{ModelRHPpics}. The RHP then reads: $\widehat{\bf \Psi}$ is analytic
in $\mathbb{C}\backslash C$ with $C = C_1 \cup C_2 \cup C_3 \cup C_4$ with the jump conditions
\begin{align}
\widehat{\bf \Psi}_+ (z) = \widehat{\bf \Psi}_- (z) \widehat{\bit{M}}
\, ,
\end{align}
and the jump matrices being
\begin{align}
\widehat{{\bit M}}_{C_1, C_3} = 
\left(
\begin{array}{cc}
1 & 0 \\
1 & 1
\end{array}
\right)
\, ,
\qquad
\widehat{{\bit M}}_{C_2} = 
\left(
\begin{array}{cc}
0 & 1 \\
-1 & 0
\end{array}
\right)
\, ,
\qquad
\widehat{{\bit M}}_{C_4} = 
\left(
\begin{array}{cc}
1 & 1 \\
0 & 1
\end{array}
\right)
\, .
\end{align}
$\widehat{\Psi}$ has the same asymptotic behavior as ${\Psi}$ in Eq.\ \re{BesselModelAsyBevavior}. Then the solution to this modified Bessel model RHP is obtained from the previous one
by the substitution $H (0,z) \to H(z_0, z)$, i.e., \cite{ChaDoe2017}.
\begin{align}
\label{ModBesModelRHP}
\widehat{{\bit \Psi}} (z)
&
=
\left(
\begin{array}{cc}
I_0 (\sqrt{z})  
& 
\ft{i}{\pi} K_0 (\sqrt{z})  
\\
i \pi \sqrt{z} I_1 (\sqrt{z})  
& 
\sqrt{z} K_1 (\sqrt{z})  
\end{array}
\right)
H (z_0, z)
\, , 
\end{align}
and the same assignment of $H$ in the three regions as above (of course, ${\rm II} = \widehat{\rm II}$, ${\rm III} = \widehat{\rm III}$  and ${\rm I} = \widehat{\rm I} \cup \widehat{\rm IV}$).

{\bf B. Airy model RHP.} The solution to the Airy model RHP is analytic everywhere in the complex plane except on the contour 
$C = C_1 \cup C_2 \cup C_3 \cup C_4$ shown in the  panel $(c)$ of Fig.\ \ref{ModelRHPpics}, where it develops jumps
\begin{align}
\bit{\phi}_+ = \bit{\phi}_- \widehat{\bit M}
\end{align}
with matrices
\begin{align}
\widehat{\bit M}_{C_1, C_3} = 
\left(
\begin{array}{cc}
1 & 0 \\
1 & 1
\end{array}
\right)
\, ,
\qquad
\widehat{\bit M}_{C_2} = 
\left(
\begin{array}{cc}
0 & 1 \\
-1 & 0
\end{array}
\right)
\, ,
\qquad
\widehat{\bit M}_{C_4} = 
\left(
\begin{array}{cc}
1 & 1 \\
0 & 1
\end{array}
\right)
\, ,
\end{align}
and the asymptotics
\begin{align}
\label{AiryModelAsy}
\bit{\phi}(z) \stackrel{z \to \infty}{\to} z^{- \sigma_3/4} A \e^{- \frac{2}{3} z^{3/2} \sigma_3}
\, .
\end{align}
The model was introduced in Ref.\ \cite{AiryModelRHP} and the solution is
\begin{align}
\label{AirySol1}
{\bit\phi}_{\rm I} (z)
&
=
\sqrt{2 \pi}
\left(
\begin{array}{cc}
{\rm Ai} (z) & - \omega^2 {\rm Ai} (\omega^2 z) \\
-i {\rm Ai}^\prime (z) & i  \omega {\rm Ai}^\prime (\omega^2 z)
\end{array}
\right)
\, , \\
{\bit\phi}_{\rm II} (z)
&
=
{\bit\phi}_{\rm I} (z)
\left(
\begin{array}{cc}
1 & 0 \\
-1 & 1
\end{array}
\right)
\, , \\
{\bit\phi}_{\rm III} (z)
&
=
{\bit\phi}_{\rm IV} (z)
\left(
\begin{array}{cc}
1 & 0 \\
1 & 1
\end{array}
\right)
\, , \\
\label{AirySol4}
{\bit\phi}_{\rm IV} (z)
&
=
\sqrt{2 \pi}
\left(
\begin{array}{cc}
{\rm Ai} (z) & \omega {\rm Ai} (\omega z) \\
- i {\rm Ai}^\prime (z) & - i \omega^2 {\rm Ai}^\prime (\omega z)
\end{array}
\right)
\, ,
\end{align}
where $\omega = \e^{2 \pi i /3}$.


\end{document}